# Approaching the limits of optoelectronic performance in mixed cation mixed halide perovskites by controlling surface recombination

*Sarthak Jariwala[1,2], Sven Burke[1,2], Sean Dunfield[3,4], Clayton Shallcross[5], Margherita Taddei[1], Jian Wang,[1] Giles E. Eperon[3,6], Neal R. Armstrong[5], Joseph J. Berry[3], David S. Ginger[1]\**

1. Department of Chemistry, University of Washington, Seattle, WA 98195, USA
2. Department of Materials Science and Engineering, University of Washington, Seattle, WA 98195, USA
3. National Renewable Energy Laboratory, Golden, CO 80401, USA
4. Materials Science & Engineering Program, University of Colorado Boulder, Boulder, CO 80309
5. Department of Chemistry and Biochemistry, University of Arizona, Tucson, Arizona 85721, USA
6. Swift Solar Inc., San Carlos, CA 94070, USA

*Corresponding author: dginger@uw.edu

Abstract:

We demonstrate the critical role of surface recombination in mixed-cation, mixed-halide perovskite, $FA_{0.83}Cs_{0.17}Pb(I_{0.85}Br_{0.15})_3$. By passivating non-radiative defects with the polymerizable Lewis base (3-aminopropyl)trimethoxysilane (APTMS) we transform these thin films. We demonstrate average minority carrier lifetimes > 4 $\mu$s, nearly single exponential monomolecular PL decays, and concomitantly high external photoluminescence quantum efficiencies (>20%, corresponding to ~97% of the maximum theoretical quasi-Fermi-level splitting) at low excitation fluence. We confirm both the composition and valence band edge position of the $FA_{0.83}Cs_{0.17}Pb(I_{0.85}Br_{0.15})_3$ perovskite using multi-institution, cross-validated, XPS and UPS measurements. We extend the APTMS surface passivation to higher bandgap double cation (FA,Cs) compositions (1.7 eV, 1.75 eV and 1.8 eV) as well as the widely used triple cation (FA,MA,Cs) composition and observe significant PL and PL lifetime improvements after surface passivation. Finally, we demonstrate that the average surface recombination velocity (SRV) decreases from ~1000 cm/s to ~10 cm/s post APTMS passivation for $FA_{0.83}Cs_{0.17}Pb(I_{0.85}Br_{0.15})_3$. Our results demonstrate that surface-mediated recombination is the primary non-radiative loss pathway in MA-free mixed-cation mixed-halide films with a range of different bandgaps, which is a problem observed for a wide range of perovskite active layers and reactive electrical contacts. This study indicates that surface passivation and contact engineering will enable near-theoretical device efficiencies with these materials.

Over the last decade, halide perovskites have demonstrated rapid improvements in power conversion efficiency (PCE), with the current single-junction PCE record at 25.2% and the perovskite/Si 2-terminal tandem at 29.2%.[1] Much of this improvement in materials performance can be attributed to cation and anion engineering[2–11] on the A and X site of the $ABX_3$ perovskite structure, respectively. However, despite the wide range of cation combinations available, the majority of the best performing single-junction perovskite solar cells, as well as high-performing tandem solar cells reported to date, incorporate some fraction of methylammonium (MA) in the A-site.[2,3,5,6,11–15] While MA-incorporation may help with performance, it has also been linked to instability at elevated temperatures achieved during regular solar cell operation due to methylamine sublimation.[16,17] Although new encapsulation strategies can enable stable MA incorporated devices,[18–20] few high-performing devices in the literature have explored completely MA-free compositions, though there are increasing efforts focusing in that direction.[8,21]

Regardless of the perovskite composition, non-radiative recombination occurring at the surfaces and interfaces (both grain boundaries and the contact layer/perovskite interface) is one of the limiting factors for achieving near-theoretical efficiencies.[22–29] Surface passivation[15,22,25,30–32] and interfacial surface and strain engineering[13,14,23,33–35] have been proposed as ways to reduce non-radiative recombination. The underlying philosophy behind these passivation schemes stems from the fact that, for the above mentioned 3D perovskites termination of the lattice along any low index face, at the surface of the crystal, at interfaces with electrical contacts, or at misaligned crystal (grain) boundaries, exposes under-coordinated metal sites, halide vacancies, and in general, regions which depart from the bulk stoichiometry and which are energetically distinct (providing potential mid-gap states) and potential recombination sites. Lewis bases of varying strengths have appeared to provide for some degree of mitigation and passivation of these interfacial states.[15,26,30,31]

For instance, surface passivation with TOPO on the archetypal $CH_3NH_3PbI_3$ has demonstrated quasi-Fermi level splitting (QFLS) >97% of the radiative limit, rivaling that of GaAs,[25] and surface recombination velocities (SRV) <10 cm/s.[22] This suggests that non-radiative recombination at the surfaces can be mitigated with appropriate surface passivation. However, we have found that TOPO passivation is not as effective for FA-containing perovskites as it is for $CH_3NH_3PbI_3$ (Figure S1, S2), as have others (private communication with Dr. Michael McGehee). Moreover, as shown previously, TOPO and other surface-passivating small molecules such as ODT (octadecanethiol) are labile and can be removed from the surface after deposition.[30]

Consequently, there is a need to develop other passivation strategies that work well across perovskite compositions and are also more compatible with solar cell processing at scale. Ideally, these strategies should target molecules which simultaneously: (1) passivate surface defects, enabling ideal radiative performance of the semiconductor; (2) remain stable to subsequent vacuum or solution processing (perhaps by crosslinking); (3) are in principle scalable; and (4) permit effective charge extraction. However, since there is limited proof-of-concept for even meeting criteria (1), (2), and (3) simultaneously, we here address meeting those criteria. Specifically, herein we explore the use of the cross-linkable[36–38] Lewis base (3-aminopropyl) trimethoxy silane (APTMS) to passivate 4 different FA-Cs/I-Br Pb perovskites with bandgaps ranging from 1.63 to 1.80 eV as well as the widely used triple cation perovskite $(FA_{0.83}MA_{0.17})_{0.95}Cs_{0.05}Pb(I_{0.83}Br_{0.17})_3$. We demonstrate an enhancement in photoluminescence quantum efficiency (PLQE), photoluminescence (PL) lifetimes, and concomitantly low SRV at interfaces, indicating the ability of this agent to passivate defects in a range of perovskite compositions. Moreover, for the case of 1.63 eV-bandgap FA-Cs/I-Br perovskite, we demonstrate a ~10x average PL lifetime improvement and a champion PL lifetime of 4.3 µs. Together with the increase in PL lifetime, we also observe an increase in the external PLQE to a record 20.1% at 70 mW/cm$^2$ for APTMS surface passivated films. This corresponds to ~97% of the theoretical Shockley-Queisser (SQ) limit. We note that such high PLQE and QFLS have not previously been reported in MA-free mixed-halide 3-D perovskites. Furthermore, we use X-ray photoelectron spectroscopy (XPS) and glow discharge optical emission spectroscopy (GDOES) to demonstrate that APTMS interacts as a surface modifier with terminal amines acting as a Lewis base. Lastly, we investigate the impact of surface passivation on the SRV in FA-Cs/I-Br films. We demonstrate that APTMS-passivated films can achieve SRVs <10 cm/s compared to control films with ~1000 cm/s. This decrease in SRV should translate to an absolute increase of >4% in PCE for surface passivated films according to our drift-diffusion simulations. Our results demonstrate that non-MA based compositions can achieve high performance by tailoring the surface chemistry to minimize non-radiative recombination occurring at the interfaces, suggesting that proper surface passivation and interface engineering will enable these compositions to be regularly incorporated in high performing devices.

To start, we fabricate $FA_{0.83}Cs_{0.17}Pb(I_{0.85}Br_{0.15})_3$ (hereafter referred to as Cs17Br15 for brevity) on glass using a method adopted from Kim *et al*. (see SI for more details).[21] Then, we confirm the perovskite structure using X-ray diffraction (XRD) (Figure S3) and bandgap of 1.63 eV using UV-Vis (Figure S4). Moreover, using XPS we show that the near surface film composition is identical (within error margins) to the solution stoichiometry with a slight (±2%) excess of PbI$_2$ at the surface (Table S1). Additionally, using UV photoelectron spectroscopy (UPS) and XPS, we determine the ionization potential (IP) for the Cs17Br15 composition to be 5.64 eV, from which we extrapolate the electron affinity (EA) of 4.01 eV (Figure 1a). Importantly, we also verified that these surface composition and band edge positions are reproducible both after shipping, and upon fabricating devices at multiple institutions, providing cross validation of sample shipping, fabrication, and XPS/UPS measurement protocols (see SI for more details on multi-institutional cross verification and shipping).

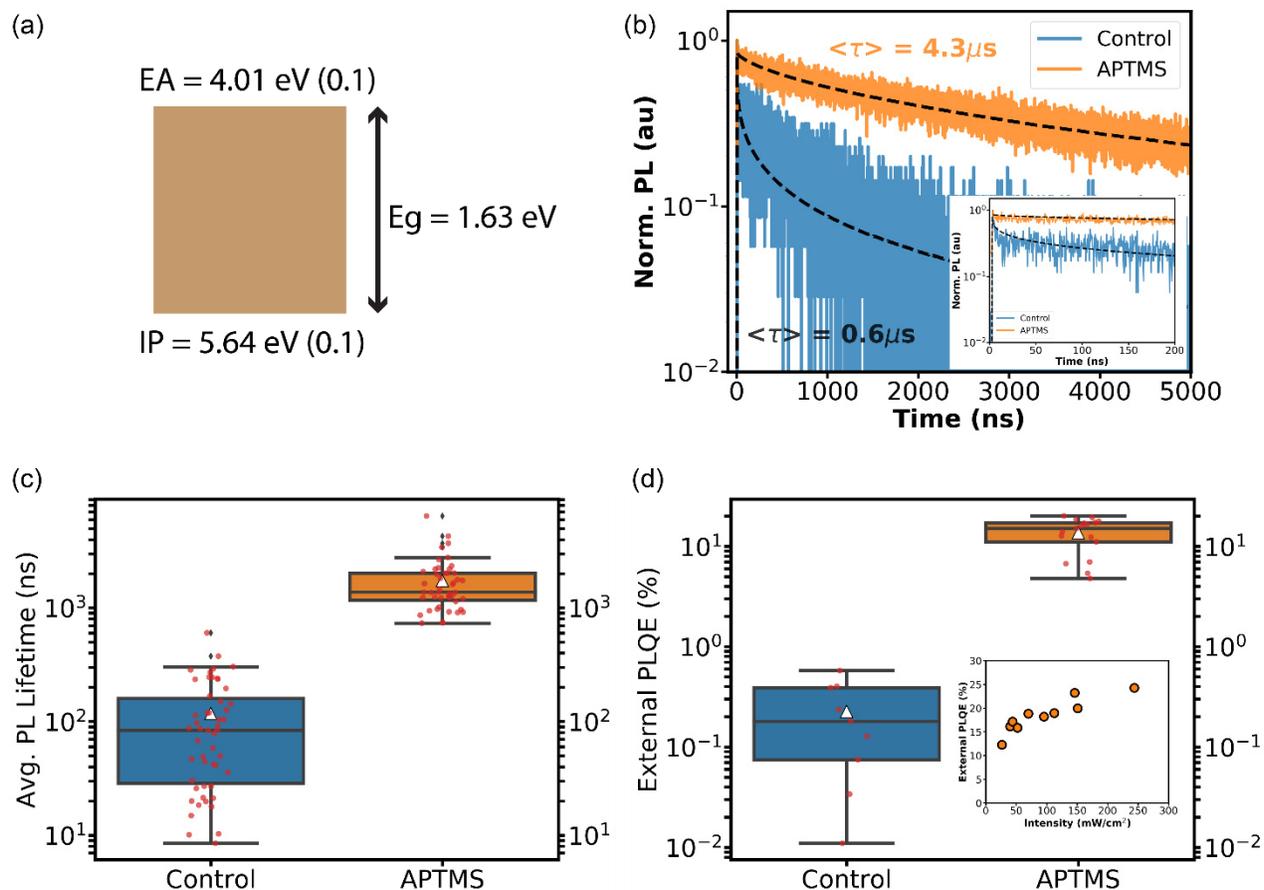

**Figure 1 – Surface passivation beyond MA based halide perovskites. (a) Valence band maximum and conduction band minimum for Cs17Br15 perovskite showing electron affinity (EA) and ionization potential (IP). (b) Time-resolved PL decay of a control and APTMS surface-passivated Cs17Br15 perovskite film. The dashed black lines denote the fitted PL lifetime according to Equation 1. The inset shows the early time decays for the control and APTMS passivated film. (c) Boxplot of average PL lifetime for control (n=51) and APTMS passivated Cs17Br15 films (n=47). (d) Boxplot of external PLQE for control (n=9) and APTMS passivated Cs17Br15 films (n=17) measured at 70 mW/cm² using a 532 nm CW laser. The inset shows the external PLQE measured as a function of intensity for a typical APTMS passivated Cs17Br15 film. The boxes in (c) and (d) shows the quartiles of the data and the whiskers extend to capture the rest of the distribution. The white triangle denotes the average of the distribution in both (c) and (d). The red scatter points are the individual data points.**

In order to passivate the surface of the Cs17Br15 films, we expose them to APTMS vapor in a controlled environment (see SI) and study the impact on PL and PL lifetimes. Figure 1b shows the PL decay of a control and APTMS surface passivated Cs17Br15 film acquired using a 470 nm pulsed laser excitation at 50 nJ/cm² per pulse (a relatively low fluence, where the trap density is generally higher than the generated carrier density[27,39]). To analyze the data, we fit the time resolved PL decays to the stretched exponential decay function described in Eq 1. Physically, the stretched exponential can be interpreted as a superposition of different relaxation rates in the film and captures the distribution of local recombination rates.[23,27,30,40] The distribution of the relaxation rates is captured by the β factor in Eq 1. A β value closer to 0 represents a more heterogeneous distribution in the relaxation rates and a β value closer to 1 represents a more homogeneous distribution in the relaxation rates. If β is 1, then the PL decay is a single exponential. The characteristic lifetime ($\tau_c$ in Eq 1) is the time required for the PL intensity to reach 1/e of the maximum intensity. The average lifetime, or <τ>, for a single PL decay considers the characteristic lifetime and the β factor to give an average lifetime of that distribution, and is given by Eq 2. Notably <τ> is the lifetime most relevant for analyzing photophysics, such as evaluating PL quantum yields.[41]

$$I = I_0 e^{(-t/\tau_c)^\beta} \quad (1)$$

$$<\tau> = \frac{\tau_c}{\beta} \Gamma(1/\beta), \text{ where } \Gamma\left(1/\beta\right) \text{ is the gamma function} \quad (2)$$

Before surface passivation, the PL lifetimes (average <τ>) of the control samples are hundreds of ns (control$_{avg}$ = 117 ns), indicating the good quality of our control Cs17Br15 thin films.[21] After APTMS surface passivation, we observe a ~15x improvement in PL lifetime (APTMS$_{avg}$ = 1.75 μs), as shown by the average lifetime box plots for 50 separate Cs17Br15 films from different batches in Figure 1c. We observe an improvement in average PL lifetime (<τ>) from 0.6 μs for the champion control Cs17Br15 film to 4.3 μs for the champion APTMS passivated film. As expected, the characteristic lifetime ($\tau_c$) also increases from 0.056 μs for the control to 2.95 μs for the surface passivated film and β also increases from 0.29 to 0.62. These data show the APTMS surface treatment modifies surface mediated non-radiative recombination and also suggest that the relaxation rate becomes more homogeneous with APTMS passivation. Concomitant with the increase in the PL lifetimes, we also observe an increase in the external PLQE with APTMS passivation. Figure 1d shows the external PLQE for films before and after passivation measured at an incident excitation intensity of 70 mW/cm² using a 532 nm CW laser (see SI for details). On average, we observe a ~60x increase in the external PLQE from 0.23% for the control to 13.64% for the passivated film, with the highest external PLQE of 20.1% at 70 mW/cm$^2$ for APTMS passivated Cs17Br15 films. We note that, to our knowledge, this is the highest reported external PLQE for MA-free mixed-cation mixed-halide perovskites. Additionally, external PLQE can be used to evaluate QFLS within the film according to the equation by Ross,[42] and our group[25] and others[43–45] have used it to evaluate QFLS in semiconductor thin films. More specifically, we have previously independently verified that the QFLS value obtained from the Ross equation[42] is the same as the QFLS obtained from fitting the full calibrated absolute intensity photoluminescence spectrum to the occupation-density-of-states corrected Planck equation.[25] Here, using the Ross equation, we calculate a 1.31 eV QFLS, i.e. ~97% of the theoretical SQ limit (see SI), obtained at 70 mW/cm$^2$ (~1 sun condition) for APTMS passivated Cs17Br15 (see SI for details). This result further demonstrates that with surface passivation, FA-Cs/I-Br based perovskites can achieve similar thin-film optoelectronic qualities without the need for MA inclusion.[25,26] These results also then indicate that the use of MA may impact the surface defect density and thus, the overall nonradiative recombination rate for typical thin-film processing routes.

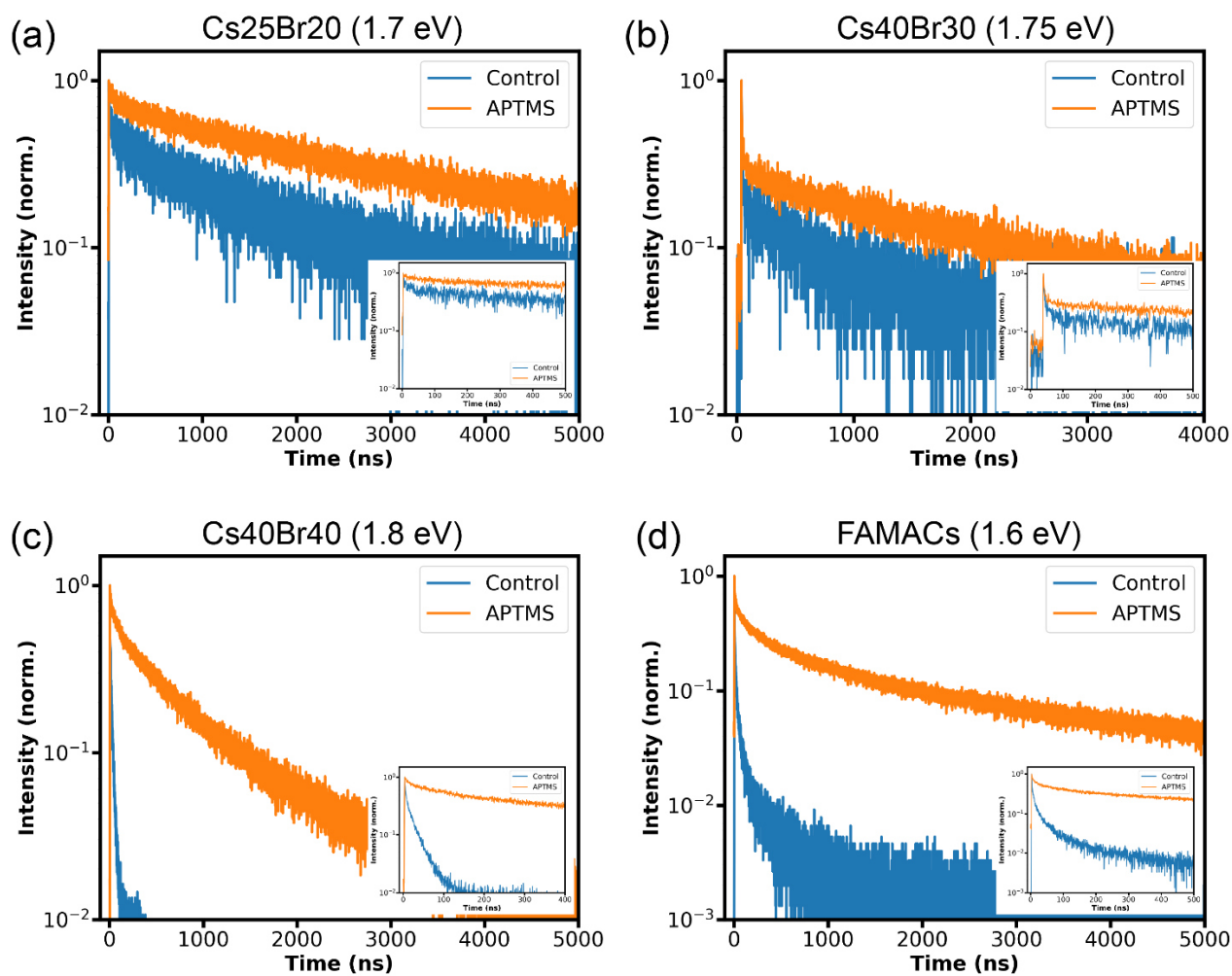

**Fig 2- Surface passivation for different bandgap MA-free perovskite compositions and triple cation compositions.** Time-resolved PL decay of control and APTMS passivated (a) Cs25Br20 (1.7 eV), (b) Cs40Br30 (1.75 eV), (c) Cs40Br40 (1.8 eV), and (d) $(FA_{0.83}MA_{0.17})_{0.95}Cs_{0.05}Pb(I_{0.83}Br_{0.17})_3$ or FAMACs-IBr (1.6 eV) perovskite thin films.

Next, we investigate the effects of surface passivation on the widely used triple-cation mixed-halide perovskite, $(FA_{0.83}MA_{0.17})_{0.95}Cs_{0.05}Pb(I_{0.83}Br_{0.17})_3$, (hereafter FAMACs/I-Br)[2] and other higher bandgap perovskite thin films (without MA) potentially relevant for Si/perovskite tandems and perovskite/perovskite tandems (1.7 eV, 1.75 eV, 1.81 eV).[46] We tune the Cs and Br composition to achieve relevant bandgaps using a method adopted from Bush *et al.*[47] (see SI for fabrication details). More specifically, we fabricate $FA_{0.75}Cs_{0.25}Pb(I_{0.80}Br_{0.20})_3$ [Cs25Br20] ($E_g$ ~1.7 eV), $FA_{0.60}Cs_{0.40}Pb(I_{0.70}Br_{0.30})_3$ [Cs40Br30] ($E_g$ ~1.75 eV), and $FA_{0.60}Cs_{0.40}Pb(I_{0.60}Br_{0.40})_3$ [Cs40Br40] ($E_g$~1.81eV) thin films and confirm the bandgap and structure of the fabricated films using UV-Vis (Figure S4) and XRD (Figure S3), respectively. Figures 2 a, b, and c show the PL decay before passivation (control) and after APTMS passivation for Cs25Br20, Cs40Br30, and Cs40Br40 films, respectively. Figure 2d shows the PL decay for the FAMACs/I-Br before and after surface passivation. We observe an increase in the PL lifetime and the integrated PL intensity (Figure S5, S6) upon APTMS treatment for all the different bandgaps and compositions. Moreover, this increase in PL lifetime tends to track linearly with the increase in silane deposition time (Figure S7) and can also be achieved through solution processing (Figure S8). This further demonstrates the versatility of the silane deposition for passivation.

We have, therefore, demonstrated here that surfaces of MA-free mixed-cation mixed-halide perovskites introduce non-radiative recombination that can be passivated using a suitable Lewis base regardless of the bandgap or the composition of the perovskite.

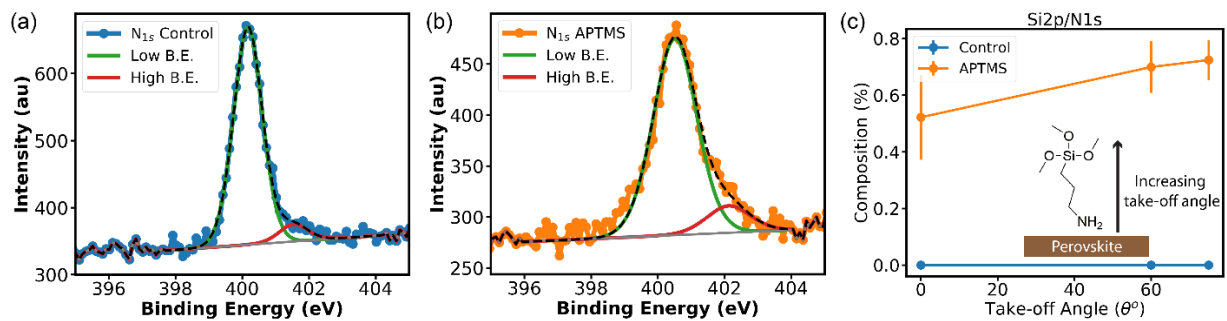

**Figure 3 – High resolution and angle-resolved X-ray photoelectron spectroscopy (XPS). High-resolution XPS of N 1s for (a) Cs17Br15 control and (b) APTMS passivated film. The dashed black line represents the envelope of the signal and the gray line represents the background. (c) $Si_{2p}/N_{1s}$ ratio as a function of different photoelectron take-off angle for control and APTMS**

**passivated Cs17Br15 films. Increasing take-off angles represent more surface sensitivity. Error bars represent the standard deviation of the composition. The inset shows the schematic of APTMS orientation on perovskite surface.**

In order to understand the general interaction between APTMS/silane and the perovskite surface, we use X-ray photoelectron spectroscopy (XPS). XPS provides information about the elements present, and their chemical state, simultaneously. The XPS results, show an increase in the Si and O concentrations after APTMS passivation of Cs17Br15 films (Figure S9), consistent with the presence of APTMS on the surface. Additional elemental depth profiles acquired using glow discharge optical emission spectroscopy (GDOES) also indicate the presence of APTMS on the surface and further demonstrate that APTMS only acts as a surface modifier and does not permeate the film at detectable levels (Figure S10). Moreover, SEM images (Figure S11) show that the general morphology of the perovskite remains largely unchanged. However, it is noteworthy that the silane layer does tend to reduce the SEM image clarity, so we cannot definitively exclude the possibility that some APTMS interacts preferentially with the domain/grain boundaries. Indeed, it is possible that exposure to the APTMS may help "clean out" grain boundaries, as has been proposed by Shallcross *et al.* during experiments on electrochemical etching of halide perovskites,[48] thus alleviating non-radiative recombination occurring at the grain boundaries. Nevertheless, the effect, dramatically increasing the PL lifetime and PLQY, is consistent with reduction of non-radiative recombination.

To further probe the interaction between APTMS and the perovskite, we acquire high-resolution XPS spectra for APTMS-passivated and control Cs17Br15 films. Figure 3a and 3b shows the spectra of N1s for the control and APTMS treated film, fitted with a low binding energy (LBE) and a high binding energy (HBE) N 1s peak. Both peak width broadening, and position shift are observed after the APTMS treatment, indicating that the N chemical bonding environment has changed. We attribute the control sample LBE and HBE N 1s peak to C-NH$_2$ and C=NH$_2^+$, respectively.[49] On the other hand, the N 1s signal in silane-treated surfaces can be assigned to a variety of overlapping amine/ammonium species interactions.[50] Here, in APTMS passivated films, we attribute the LBE N 1s peak to the free amine and H-bond donor and the HBE N1s to a range of different species such as Lewis base, H-bond acceptor, and protonated amine; as suggested by Shallcross *et al*.[50] The binding energies are summarized in Table S2. We observe a higher fraction of the electron deficient HBE N 1s in APTMS treated sample (12.15%) compared to control sample (7.32%). In other words, there is 1.6x more HBE N 1s contribution to the overall N 1s signal for APTMS treated films, compared to the control N 1s signal. We attribute this increase in the contribution of the HBE N 1s peak to the APTMS treatment acting as a Lewis base on Cs17Br15 films, as also observed in other silane/metal interactions.[50]

Next, to further understand the interaction between the APTMS and the surface of the perovskite film, we use angle resolved XPS to investigate the composition as a function of depth, allowing us to discern the average orientation of the silanes on the surface of the perovskite film. In this technique, the elemental composition is measured as a function of different photo-electron take-off angles, allowing depth resolution. Higher take-off angles represent more surface sensitivity (probing decreased film depth) and lower take-off angles represent more bulk sensitivity (probing increased film depth). Figure 3c shows the compositional ratio of Si$_{2p}$/N$_{1s}$ as a function of take-off angle for APTMS passivated Cs17Br15 films. We observe an increase in the Si$_{2p}$/N$_{1s}$ signal as a function of increasing take-off angle (decreasing film depth). In other words, we probe more $Si_{2p}$ signals, as opposed to $N_{1s}$, at the very surface of the APTMS passivated Cs17Br15 films. As discussed above, the $N_{1s}$ signal after APTMS passivation originates mainly from the amine-perovskite interactions (Figure 3b), rather than from the FA species in the bulk perovskite (Figure 3a), such a $Si_{2p}/N_{1s}$ depth profile indicates that the overall silane orientation is more upright with the terminal amines in the silane interacting with the defects at the surface (Figure 3b inset). In addition, as we move to grazing angles (decreasing film depth, increasing take-off angle), we observe a decrease in the $N_{1s}/C_{1s}$ signal (Figure S12). Together, these N, C, and Si trends suggest both that the silane films are relatively compact, and that the *average* silane orientation is more upright (Figure 3c inset) with the terminal amines in the silane pointing downward (*i.e. interacting with the defects at the surface*). We note that while it is also possible that the O species in the methoxy groups also interact with the uncoordinated Pb, the data suggest the average interaction is more N dominated (Figure S13). Such equivocality in silane arrangements have also been observed in more-studied silane deposition on Si and oxide surfaces.[51] We note that the N, C, Si and O trends discussed also hold for all different Cs17Br15 samples passivated with silanes (Figure S12-S14). Moreover, we do not observe any specific trends in N 1s or C 1s signal for control films as a function of take-off angles (different surface depth sensitivity) (Figure S14). Together, the high-resolution and angle resolved XPS results confirm that the terminal amine primarily interacts with the perovskite surface by acting as a Lewis base and donating electron density to passivate the defects, most likely at under-coordinated $Pb^{2+}$ sites at the perovskite surface.

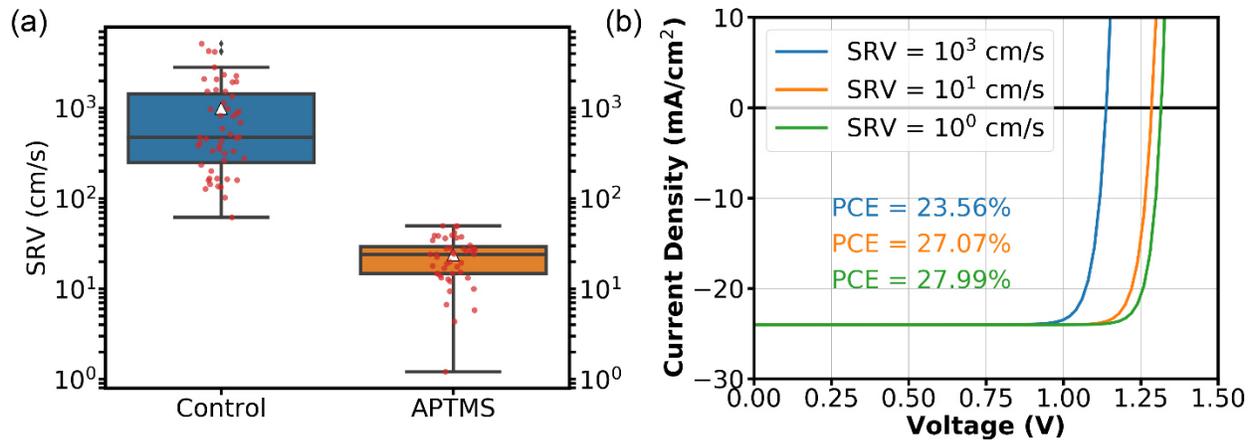

**Figure 4** – Surface recombination velocity (SRV) and its impact on $V_{OC}$ and PCE for Cs17Br15 films. (a) Boxplot of SRV for control and APTMS passivated Cs17Br15, for SRV1=0 limit. The box shows the quartiles of the data and the whiskers extend to capture the rest of the distribution. The white triangle denotes the average of the distribution. The red scatter points are the individual data points. (b) Simulated J-V curve (using SCAPS[52]) for metal/HTL/Cs17Br15/ETL/metal geometry with the same SRV at the perovskite/HTL and perovskite/ETL interface and perfect interface energy alignment of the transport layers with Cs17Br15 (HTL – hole transport layer, ETL – electron transport layer).

Having demonstrated a method to alleviate the non-radiative recombination introduced by surfaces in non-MA based mixed halide perovskites, we turn to investigate its impact on SRV. SRV quantifies the rate of minority carrier recombination occurring at the surface of the semiconductor and is one of the parameters that defines non-radiative recombination occurring at the surfaces and interfaces in devices. For the purposes of our SRV discussion here, "surfaces" will refer to the top and bottom film surfaces, though we cannot exclude the possibility that low concentrations of APTMS are also penetrating the grain boundaries, which have also been shown to act as non-radiative recombination sites.[27,40] Thus, reducing SRV has proved to be an important parameter to improve the efficiency of established PV systems including Si,[53,54] GaAs,[55] CIGS[56] and CdTe.[57]

Early modeling, using literature-reported values for the physical properties of $CH_3NH_3PbI_3$, has shown that SRVs of 10 cm/s or less will be required to approach the SQ performance limit for perovskite PV.[22] While passivated and well-grown MA-based perovskite thin-films and single crystals have already been demonstrated with relatively low SRVs on the order of ~10 cm/s, [22,58,59] unpassivated $MAPbBr_3$ single crystals,[60] $MAPbI_3$ single crystals and thin films[61] commonly report much higher SRVs on the order of ~$10^3$ cm/s.

Here, we use time-resolved PL measurements to investigate the minority carrier lifetime and the associated SRV in Cs17Br15 films. As we note above, the average PL lifetime ($<\tau>$) as given by Eq. 2 for Cs17Br15 films is the effective minority carrier lifetime under low fluence. The relationship between effective lifetime ($\tau_e$), bulk lifetime ($\tau_b$), and surface lifetime ($\tau_s$)[62] is shown in SI. The bulk lifetime ($\tau_b$) can be measured either in a single crystal or, in a case where all surface recombination is alleviated, and only bulk recombination remains (such that the internal PLQY approaches the radiative limit). Here, we use the latter approach and approximate the bulk lifetime to be 8 µs.[22,25] We note that this is an estimate of the bulk lifetime as it is measured in surface passivated thin films which have internal PLQY greater than 90% and very small amounts of bulk non-radiative recombination left.[25] Further, this is a conservative approach, as underestimating the bulk recombination rate (*i.e.* overestimating bulk lifetime) should lead to an overestimate of the extracted SRV. Using this bulk lifetime, we then use the approach described previously[22] (see SI for full details) to estimate the SRV. Our results show that on average, control films exhibit SRV of ~1000 cm/s while the same films, after surface passivation with APTMS, exhibit an average SRV of ~10 cm/s, with a champion SRV value of 1 cm/s (see SI). We emphasize that under this approach, the SRV values we report are on the *higher* end of the limit as bulk lifetimes are likely overestimated and all recombination is assumed to occur at only one interface; *i.e.* this assumption ignores recombination occurring at the back interface.

Next, we use the calculated SRV values and idealized contacts with realistic transport properties to simulate the device performance using drift-diffusion simulations in SCAPS.[52] By idealized contacts with realistic transport properties, we mean that we take the electronic transport properties (e.g. carrier mobility, thickness) typical of HTL and ETL layers, and assume that versions with appropriate energy level offsets (0 eV) and SRV (as modeled) can be realized (see SI for more details). Keeping an ideal alignment of all the contact layers allows us to explicitly simulate and understand the impact of SRV. The SRV is introduced both between the perovskite/HTL and perovskite/ETL interfaces. Although these two interfaces will likely have different SRVs in reality,[22] for the purposes of the simulation we kept their SRV values the same. Figure 5b shows the simulated J-V curve for SRV of $10^3$ cm/s, 10 cm/s and 1 cm/s. We observe a >4% absolute improvement in PCE as we vary the SRV from $10^3$ cm/s down to 1 cm/s, demonstrating the importance of minimizing SRV at the perovskite/transport layer interfaces. As expected, the majority of the improvement in PCE is due to the improvement in the $V_{OC}$ and fill factor (Figure S15). We note that the improvement in the simulated device performance due to a reduction in SRV is seen despite perfect energetic alignment of contact layers with the perovskite. Increasing the energy mismatch in transport layer alignment with the perovskite will adversely impact the PCE and $V_{OC}$ (Figure S16) placing even more stringent demands on SRV. We note that the increased importance of SRV reported here (relative to earlier work by Wang *et*

*al.*[22]) is due to our use of more realistic transport-layer properties such as mobility, etc. (see SI for a complete list of SCAPS simulation parameters). These results only increase the importance of further reducing SRV.

In conclusion, we have demonstrated high performing MA-free mixed cation and mixed halide perovskites with >4 µs PL lifetime and >20% external PLQE achieved by facile surface passivation using a silane. The high PLQE observed corresponds to a ~97% of the theoretical QFLS; a first for non-MA based compositions. Importantly, we also demonstrate that the passivation approach works for a wide range of different compositions with different bandgaps, including the widely used triple cation composition, $(FA_{0.83}MA_{0.17})_{0.95}Cs_{0.05}Pb(I_{0.83}Br_{0.17})_3$. We used XPS and UPS to confirm the stoichiometry and investigate the band alignments of the perovskite compositions. Importantly, we also demonstrate that the same values can be obtained at multiple institutions with films prepared at, and transported between, different institutions, validating protocols for sample transport and surface analysis. In addition, with high-resolution XPS and angle resolved XPS we investigate the nature of the interaction and the average orientation of APTMS on the perovskite surface. We find that, on average, the APTMS molecules primarily interact with the perovskite surface using the terminal amine group as a Lewis base. Lastly, we show the effects of APTMS passivation on the SRV and observe that the SRV decreases to ~10 cm/s on average, with a champion low of ~1 cm/s (note that this is the upper limit of the value). Using SCAPS simulations,[52,63] we demonstrate that such drastic improvements in the SRV can translate to >4% absolute enhancement in device PCE, even with perfectly aligned contacts.

Our results demonstrate that non-radiative recombination at surfaces is a major loss pathway for perovskites and by reducing the SRV with appropriate surface passivation and interface engineering, perovskites can achieve near theoretical PCE. While the use of thick APTMS in traditional n-i-p or p-i-n solar cell geometry would likely not be feasible due to its presumed insulating nature, we propose that molecules like it may still be useful in PERC-like[64] perovskite solar cells, either with self-assembled, or intentionally patterned interfacial passivation between the perovskite and the HTL and ETL. Alternatively, the use of APTMS with an interdigitated back contact perovskite solar cell[65,66] could be another practical option. Regardless, the data here provide guidance on what to look for while designing and optimizing new "passivating" layers and more importantly, they demonstrate that current perovskite thin film growth methods can already yield films with optoelectronic quality that can enable much higher potential performance than has been realized in device structures. These results thus further emphasize the importance of controlling SRV by interface passivation, even under the best of conditions such as perfectly aligned contacts.

Acknowledgements:

This material is based upon work supported primarily by the U.S. Department of Energy's Office of Energy Efficiency and Renewable Energy (EERE) under the Solar Energy Technology Office (SETO), Award Number DE-EE0008747. Part of this work was conducted at the Molecular Analysis Facility, a National Nanotechnology Coordinated Infrastructure site at the University of Washington which is supported in part by the National Science Foundation (grant NNCI-1542101), the University of Washington, the Molecular Engineering & Sciences Institute, and the Clean Energy Institute. SJ thanks Gerry Hammer at Molecular Analysis Facility for help with XPS measurements and analysis. Research support for work performed at the University of Arizona was provided by the Office of Naval Research ONR: N00014-18-1-2711 and ONR N00014-20-1-2440. Contributions from JJB and SPD were undertaken at the National Renewable Energy Laboratory, operated by Alliance for Sustainable Energy, LLC, for the U.S. Department of Energy (DOE) under Contract No. DE-AC36-08GO28308 efforts here were supported by the U.S. Department of Energy's Office of Energy Efficiency and Renewable Energy (EERE) under the Solar Energy Technologies Office (SETO) project "De-risking Halide Perovskite Solar Cells" program (DE-FOA-0000990). The views expressed in the article do not necessarily represent the views of the DOE or the U.S. Government.

Author Contributions:

SJ and SB carried out PL experiments under the supervision of DSG. SJ, SB and MT carried out PL experiments with TOPO. SPD carried out the XPS and UPS measurements at NREL under the supervision of JJB. CS carried out the XPS and UPS measurements at University of Arizona under

# Supplementary Information for

Approaching the limits of optoelectronic performance in mixed cation mixed halide perovskites by controlling surface recombination

*Sarthak Jariwala[1,2], Sven Burke[1,2], Sean Dunfield[3,4], Clayton Shallcross[5], Margherita Taddei[1], Jian Wang,[1] Giles E. Eperon[3,6], Neal R. Armstrong[5], Joseph J. Berry[3], David S. Ginger[1]\**

1. Department of Chemistry, University of Washington, Seattle, WA 98195, USA
2. Department of Materials Science and Engineering, University of Washington, Seattle, WA 98195, USA
3. National Renewable Energy Laboratory, Golden, CO 80401, USA
4. Materials Science & Engineering Program, University of Colorado Boulder, Boulder, CO 80309
5. Department of Chemistry and Biochemistry, University of Arizona, Tucson, Arizona 85721, USA
6. Swift Solar Inc., San Carlos, CA 94070, USA

*Corresponding author: dginger@uw.edu*

**Perovskite solution and thin film fabrication**

Perovskite Solution: 1M (3v% Formamide)

All precursors were purchased from Sigma, unless stated otherwise. Perovskite precursor solutions were formed by adding stoichiometric amounts of precursor salts FAI (greatcellsolar), CsI, $PbI_2$ (TCI/Sigma), $PbBr_2$ in DMF:DMSO (1300:640 for Cs17Br15 and 1600:360 for Cs25Br20, Cs40Br30 and Cs40Br40) and 60uL of Formamide to make a 2mL 1M solution.

Spin coating procedure

The glass substrates were plasma cleaned before spin coating. Inside the glovebox, the perovskite solution was filtered before deposition. The substrates were placed on the spin coater chuck and ~25 uL of the filtered perovskite solution was deposited on top. The substrate were spun at 4000 rpm for 60 seconds. When ~35 s were remaining, ~50 uL of CB was dropped from above. The films were then annealed on the hotplate at 100 C for 30 s and at 150 C for 10 mins.

Triple cation (FAMACs) films were made according to previously reported procedure.[1]

**Surface passivation with (3-Aminopropyl)trimethoxysilane or APTMS**

Surface passivation with APTMS was done at RT in vacuum oven (Precision Vacuum Oven Model 19) for 5-10 mins. 1 mL of APTMS was taken in a 4 mL vial and perovskite films were placed around the vial as shown in schematic S1. Silane deposition was performed under vacuum with the gauge pressure reading -30 In. of Hg relative to atmospheric pressure. The chamber dimensions are 30.5x20.3x20.3 cm (length x width x height) but to avoid coating the entirety of the inside chamber with silanes, we cover the films and APTMS vial with a 500 mL glass jar inside the chamber. After silane deposition, the films were sonicated in anhydrous CB/Toluene for 60 s. Alternatively, instead of sonicating, films can also be washed by dynamically dropping anhydrous CB or Toluene on top of the spinning sample (2000 rpm for 60 s). Finally, the samples were dried with $N_2$.

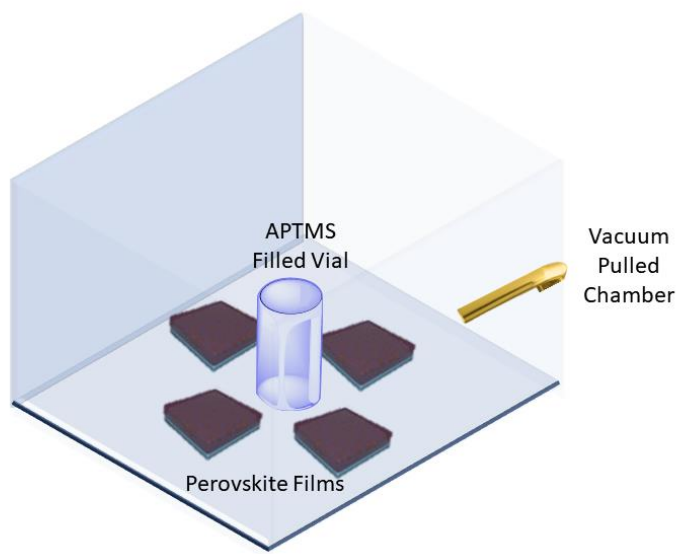

**Scheme S1. Schematic of passivation with APTMS on perovskite surfaces**

### Shipping Samples for XPS/UPS Measurements

After fabrication, the samples were transferred internally through an antechamber to a solvent-free inert glovebox. The samples were covered in an Al foil and sealed in a vacuum bag (Weston 30-0101-W 8-by-12-Inch Vacuum-Sealer Bags). The samples were stored in the solvent-free glovebox until they were shipped overnight. Upon receiving the samples, they were transferred directly to a solvent-free glovebox and stored there until XPS/UPS measurements.

### Quasi-Fermi Level Splitting (QFLS)

The Shockley-Queisser theoretical QFLS limit ($\Delta\mu_{SQ}$) for a bandgap of 1.63 eV is 1.353 eV at 300K. Using this theoretical limit in Ross equation[2], we can calculate the QFLS based on the external photoluminescence quantum efficiency (PLQE, $\eta_{ext}$) as seen in Eq. S1.

$$\Delta\mu_{\eta_{ext}} = \Delta\mu_{SQ} - kT \left|\ln(\eta_{ext})\right| \tag{S1}$$

Using measured external PLQE value of 20.1% at ~1 sun condition with above bandgap illumination and $T$ = 300 K, we calculate QFLS $\Delta\mu_{\eta_{ext}}$ value of 1.31 eV; ~97% of the theoretical SQ limit.

### External Photoluminescence Quantum Efficiency (PLQE)

PLQE was performed in an integrating sphere (Hamamatsu C9920-02, A10094) using a 532 nm CW laser (CrystaLaser, GCL532-005-L) following the procedure reported by de Mello *et al.*[3] with scattering correction. For intensity dependent measurements, the incident laser intensity was attenuated using neutral density filters

**Time-resolved Photoluminesce (TRPL)**

TRPL was measured using a PicoQuant Picoharp 300 TCSPC system equipped with a 470 nm pulsed diode laser (PDL-800 LDH-P-C-470B, 300 ps pulse width). The laser was pulsed at repetition rates from 100KHz to 1MHz. The PL emission was filtered using a 580 nm long-pass filter before being directed to the detector.

**Surface Recombination Velocity Calculation**

The surface lifetime ($\tau_s$) is determined from Eq. S2 using average PL lifetime (from PL fitting) as effective lifetime ($\tau_e$) and assuming a bulk lifetime ($\tau_b$) of 8 $\mu s$ (see main text). Using the surface lifetime, the upper limit of SRV, where one interface has no recombination (SRV1 = 0), can be calculated as shown in Eq. S3.[4] The thickness (W) can be measured experimentally (here, W = 400 nm) and the diffusion coefficient (D) can be determined using Einstein's relation (D = $\mu$kT) using measured or literature mobility ($\mu$) value of 35 cm$^2$V$^{-1}$s$^{-1}$ for Cs17Br15 thin film.[5]

$$\frac{1}{\tau_e} = \frac{1}{\tau_b} + \frac{1}{\tau_s} \tag{S2}$$

$$\tau_s = \frac{W}{SRV} + \left(\frac{4}{D}\right)(W/\pi)^2 \text{, when SRV1 = 0} \tag{S3}$$

**SCAPS Simulations**

SCAPS 3.3.07[6,7], developed at the University of Gent, was used to perform drift-diffusion simulations. Perovskite material parameters used in the simulations are listed in Table S3. We note that although photon recycling effects are not directly included in the device model, the radiative recombination rate constant is taken from experimental value measured from escaped photons post radiative recombination.[8]

The general cell design was Metal/HTL/Perovskite/ETL/Metal. In our simulations, the cell was illuminated from the HTL side. As noted in the main text, we use idealized contacts (0 eV interface energy offset) with realistic material properties such as mobility, bandgap, etc. We also keep the interface layer thicknesses the same as typical ETL, HTL layers. For a full list of the parameters used, refer to Table S3.

**Data Analysis**

The code to analyze data is available as an open-source resource at
https://github.com/SarthakJariwala/Python_GUI_apps

**Scanning Electron Microscope (SEM)**

SEM images were measured using a FEI Sirion SEM at 5 kV accelerating voltage. To avoid charging the samples were prepared on ITO substrates.

**Glow Discharge Optical Emission Spectroscopy (GDOES)**

Depth-dependent composition was measured using a Horiba GD-Profiler-2 using a high-power radio frequency (RF) argon plasma in a 4 mm diameter anode. The plasma was operated at 30 W and a pressure of 600 Pa. Lead (Pb), Bromine (Br), Iodine (I) and Oxygen (O) were detected using the 406 nm, 149 nm, 146 nm, and 130 nm atomic emission lines, respectively.

**X-ray Photoelectron Spectroscopy (XPS)**

National Renewable Energy Laboratory Measurements:

*As received samples* were loaded into a solvent free $N_2$ glovebox with < 0.1 $H_2O$ and $O_2$, unpackaged, left to sit for 30 minutes, and then loaded into a KF Flange tube that had been baked in a vacuum oven overnight at 150 C. The KF Flange tube was then shut and stored in the glovebox until XPS analysis, at which point it was moved to the XPS argon glovebox, loaded onto an analysis puck, grounded, and then transferred inertly to the analysis chamber.

*Control samples* were fabricated on ITO samples shipped to NREL using an analogous procedure to the one above. Specifically, dry salts were weighed out in a solvent free glovebox, solvated in another glovebox, and spun in a third glovebox. After annealing, the resulting perovskite films were left to sit in the glovebox for 5 minutes before being transferred to the same solvent free glovebox as the shipped samples. From this point on, samples were handled identical to the as received samples (left to sit for 30 min, loaded into dry KF, transferred to XPS glovebox for analysis).

*High resolution X-Ray Photoemission Spectroscopy (XPS)* measurements were performed on a Physical Electronics 5600 photoelectron spectrometer, which has been discussed in detail previously.[9] In order to avoid beam damage, low power I3d core level measurements were first taken with a monochromatic 15W (15kV, 1 mA) Al Kα excitation centered at 1486.7 eV, followed by typical XPS analysis using the same source at 350 W (15 KV, ~23 mA) and UPS analysis using radiation generated by a He-gas discharge lap (He I = 21.22 eV). All XPS core-level spectra were collected using a step size of 0.1 eV and pass energy of 23.50 eV while UPS spectra and XPS WF measurements were conducted with a step side of 0.025 and pass energy of 2.95 eV. The electron binding energy scale was calibrated using the Fermi edge and core levels of gold and copper substrates, cleaned with Argon ion bombardment. UPS spectra were numerically corrected for satellite peaks that arise from the polychromic He I radiation. Peak areas were fit using a Gaussian-Lorentzian peak fitting algorithm with a Shirley background. WFs were determined using the intersection between the baseline and a linear fit to the main secondary electron cutoff feature. VBMS were calculated using linear extrapolation of the main feature in the valence band edge to the background signal and corrected by the shift seen in the I3d core level measurements from low to high power. Additionally, to confirm that the consistency and reproducibility across multiple methods of analysis, fits in good agreement with the first set of measurements were made using the gaussian interpolation method described above (with different sigma factor due to different instrument response functions). Spectra taken with the Al source are typically assigned an uncertainty of 0.05 eV. Spectra taken with UPS are typically assigned an uncertainty of 0.025 eV. Compositional analyses are typically assigned an uncertainty of 5%.

University of Arizona Measurements:

*Monochromatic XPS and UPS:* Samples from the University of Washington, prepared and shipped under inert atmosphere environment, were loaded into a clean $N_2$ glovebox (< 0.1 ppm $O_2$ and $H_2O$ and solvent-free for over a week) which allowed for the samples to be positioned on a sample stub and transferred into the high vacuum surface analysis chamber without exposure to ambient. Photoemission spectroscopies were then conducted with a Kratos Axis Ultra PES system with a base pressure of ca. $2 \times 10^{-9}$ Torr. XPS core level (CL) spectra were acquired with a monochromatic Al K$\alpha$ source (1486.6 eV, 20 mA, 15 kV, 20 eV pass energy) at a take-off angle (TOA) of 0 degrees w.r.t. the surface normal. Average and standard deviation values for BEs and atomic ratios were determined by measuring at least three spots per sample (n ≥ 3). The XPS binding energy scale was calibrated with sputter-cleaned Cu ($3p_{3/2}$ = 75.10 eV; $2p_{3/2}$ = 932.60 eV), Au ($4f_{7/2}$=84.00 eV), and Ag ($3d_{5/2}$, 368.20 eV) foils. XPS spectra were processed using the Vision 2 software package (Kratos Analytical). Relative atomic concentrations are quantified from background-corrected and systematically deconvoluted CL spectra, which are corrected for the orbital photoionization cross sections and KE-dependent analyzer transfer function. Prior to typical high intensity (20 mA, 15 kV: 300 W) XPS measurements, the secondary electron cutoff (SECO) was measured using low intensity (1 mA, 15 kV: 15 W) XPS at 3 different spots on each sample, and the I 3d and Pb 4f core level (CL) were also measured at spot 2 on each sample. These measurements were used to determine if high intensity X-ray exposure changes the sample work function, band edge positions (i.e., BE with respect to a constant Fermi Energy; $BE_F$ = 0 eV) and I/Pb ratio. For purposes of this report no discernable changes were noted for these parameters although it should be noted that extensive X-ray exposure of the perovskite film under some XPS analysis conditions can cause compositional changes within the sampling depth, which can change the effective work function, and need to be monitored to ensure sample integrity is sustained.

After the XPS characterization the valence band region (high kinetic energy) and the SECO were measured at spot 2 using UPS with a He I excitation source (SPECS UVS 10/35), (ca. $1.1-1.2 \times 10^{-7}$ Torr He pressure, the He discharge lamp was operated at a current of 25 mA, producing a source energy = 21.22 eV) and an analyzer pass energy of 5 eV. Samples were fixed and grounded to stainless steel stubs. Samples were biased at -10.0 V for UPS measurements to enhance the yield of photoelectrons at the low kinetic energy (LKE) edge. The Fermi energy ($E_F$) for UPS measurements was calibrated with sputter-cleaned Au, and the work function of the sample determined by subtracting the spectral width ($E_F$-LKE) from the excitation energy (21.22 eV), providing a value of 5.1 eV for clean Au.

To ensure maximum sensitivity and spectral contrast for determination of the VBM energy and work function of each sample He $I_{\beta,\gamma}$ satellites were removed from these spectra. These UPS experiments used a non-monochromatic He discharge lamp excitation source, which has a primary He I$\alpha$ emission line at 21.22 eV. The He discharge lamp also emits low-intensity He I$\beta$ ($E_\beta$ = 23.09 eV, $\Delta E_{\beta-\alpha}$ = 1.87 eV, $I_{\beta(\alpha)}$ = 1.4-1.5% of the He I$\alpha$ intensity) and He I$\gamma$ ($E_\gamma$ = 23.74 eV,

$\Delta E_{\gamma-\alpha}$ = 2.52 eV, $I_{\gamma(\alpha)}$ = 0.6-0.8% of the He I$\alpha$ intensity) satellite lines, which generate photoelectrons with slightly higher kinetic energy (KE) that overlap with the primary He I$\alpha$ - created photoemission spectrum which can effectively hide spectral features below the VB edge, and may complicate determination of the VBM energy. In order to subtract satellite photoemission features from the He I$\alpha$ VB spectra: 1) the raw UPS data is multiplied by the relative satellite intensity ($I_{\beta/\gamma(\alpha)}$ which shows slight variations due to fluctuations in He discharge lamp pressure caused by evaporation and refilling of the liquid nitrogen trap that removes trace oxygen from the He gas source; 2) shifted by the relative offset w.r.t. He I$\alpha$ ($\Delta E_{\beta/\gamma-\alpha}$) to lower BE, and 3) subtracted from the raw UPS data. After removal of satellite photoemission features the residual background due to secondary electron photoemission was subtracted from the UPS data using a polynomial fitting function. The valence band region was then deconvoluted with the minimum number of Gaussian peaks to adequately reproduce the data, and this data was then area normalized to reveal differences in overall spectral shape. The VB onset energy was estimated using a Gaussian Interpolation protocol where VBM is determined to occur at 3.5$\sigma$ on the high kinetic energy side of the highest KE VB peak were $\sigma$ ~ the full width at half-maximum of this peak.[10] The values of VBM shown here are similar to those shown previously by Tao and Olthoff using a monochromatic He I discharge.[11]

University of Washington Measurements:

High resolution XPS and angle resolved XPS at University of Washington was performed using Kratos AXIS Ultra DLD XPS system on samples prepared on ITO substrates. Control samples were fabricated as described above. APTMS samples were treated with APTMS vapors for 5 mins. The angle-resolved composition for an individual sample was calculated by averaging the composition at minimum three different sites on the same sample at every photo-electron take-off angle. XPS spectra were analyzed using the Kratos Analytical software package.

**Table S1: Perovskite composition determined from XPS measurements**

| Made at | FA/A | Cs/A | Pb/A | (I/3)/A | (Br/3)/A | Measured |
|---|---|---|---|---|---|---|
| UW | 0.84 | 0.16 | 1.06 | 0.86 | 0.14 | Univ. of Arizona |
| UW | 0.83 | 0.17 | 1.07 | 0.86 | 0.14 | Univ. of Arizona |
| UW | 0.84 | 0.16 | 1.02 | 0.87 | 0.13 | Univ. of Arizona |
| UW | 0.84 | 0.16 | 1.09 | 0.90 | 0.16 | NREL |
| NREL | 0.84 | 0.15 | 1.08 | 0.87 | 0.15 | NREL |
| *Average* | *0.84* | *0.16* | *1.06* | *0.87* | *0.14* | |
| *Std. Dev* | *0.01* | *0.01* | *0.02* | *0.02* | *0.01* | |

**Table S2. XPS binding energies of N 1s signal for control and APTMS passivated Cs17Br15 thin films**

|  | Binding Energy | Peak Area (RSF Corrected) | % |
| --- | --- | --- | --- |
| **Control – N1s** | 400.172 eV | 7686.4 | 92.68 |
| **Control – N1s** | 401.542 eV | 606.75 | 7.32 |
| **APTMS – N1s** | 400.514 eV | 6932.63 | 87.85 |
| **APTMS – N1s** | 402.121 eV | 959.049 | 12.15 |

**Table S3. SCAPS simulation parameters.**

| Parameter | Value |
| --- | --- |
| **Perovskite Thickness (nm)** | 400 |
| **Dielectric Permittivity** | 39 |
| **Electron Affinity (eV)** | 4.01 |
| **Bandgap (eV)** | 1.63 |
| **Absorption Coefficient (cm$^{-1}$)** | SCAPS Model |
| **Bulk Lifetime (ns)** | 8000 |
| **Radiative Recombination Rate Constant (cm$^3$/s)** | 4e-11[8] |
| **Auger Rate Constant (cm$^6$/s)** | 0 |
| **VB/CB Effective Density of States (cm$^{-3}$)** | 3e18 |
| **Electron and Hole Mobility (cm$^2$/V.s)** | 35[5] |
| **Doping Concentration (cm$^{-3}$)** | 1e12 |
| **Bandgap ETL (eV)** | 2 |
| **Bandgap HTL (eV)** | 3 |
| **Electron Mobility (cm$^2$/V.s) ETL** | 1e-2 |
| **Hole Mobility (cm$^2$/V.s) HTL** | 1e-4 |
| **Minority Carrier Recombination Velocity from Perovskite to ETL/HTL (cm/s) – SRV** | Vary |
| **Thickness ETL (nm)** | 30 |
| **Thickness HTL (nm)** | 10 |
| **Interface Energy Offset ETL/HTL with Perovskite** | Vary (0 eV main text) |
| **VB/CB Effective Density of States (ETL and HTL) (cm$^{-3}$)** | 1e20 |
| **Dielectric Constant ETL** | 5 |
| **Dielectric Constant HTL** | 3.5 |
| **Majority and Minority Carrier Recombination Velocity at Front and Back Metal Contact (cm/s)** | 1e7 |
| **Interface Energy Offset ETL/HTL with Metal Contacts (eV)** | 0 |

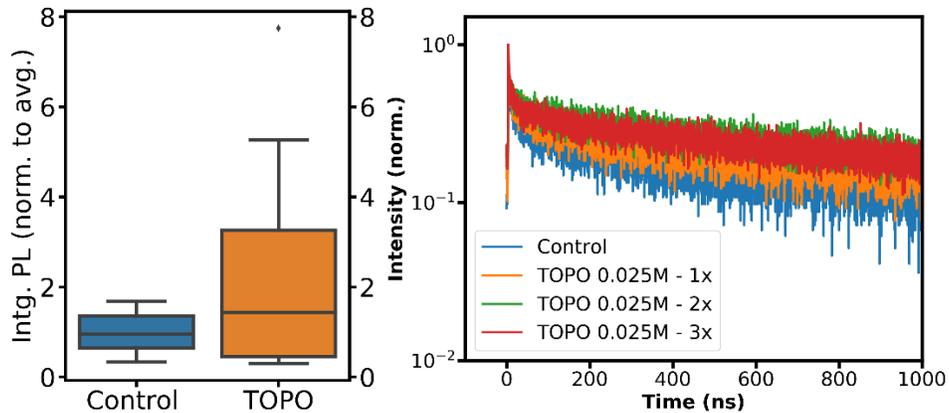

**Figure S1.** Integrated PL intensity (normalized to average) and time-resolved PL decay for control Cs17Br15 and TOPO (0.025M) treated Cs17Br15 films

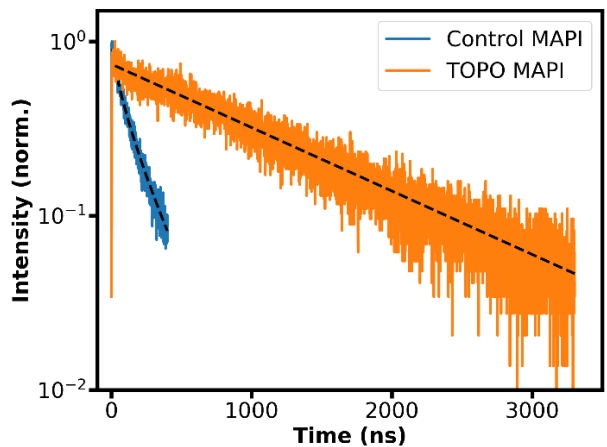

**Figure S2.** Time-resolved PL decay for CH$_3$NH$_3$PbI$_3$ (MAPI) thin films (made acc to ref.[12]) before and after TOPO treatment.[13] The average PL lifetime improves from 137 ns (control) to 1.2 $\mu$s after TOPO treatment.

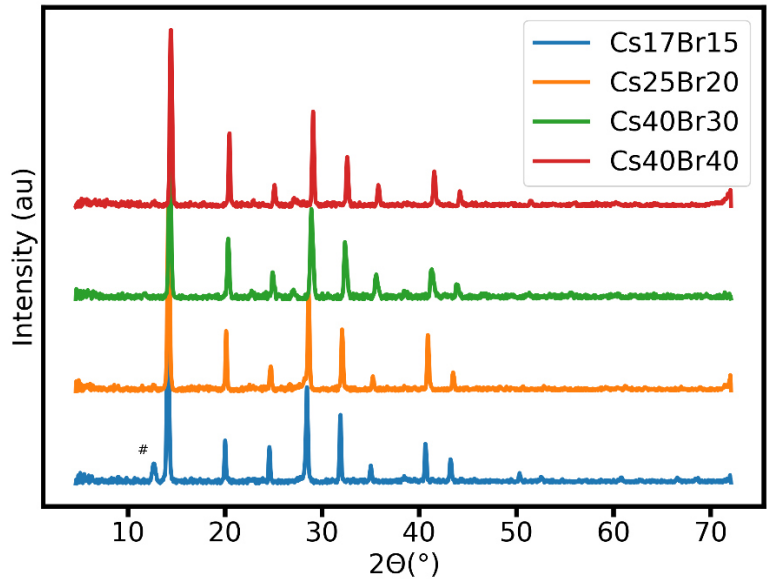

**Figure S3.** X-ray diffraction of Cs17Br15, Cs25Br20, Cs40Br30 and Cs40Br40 perovskite thin films. Peak marked with "#" represents $PbI_2$.

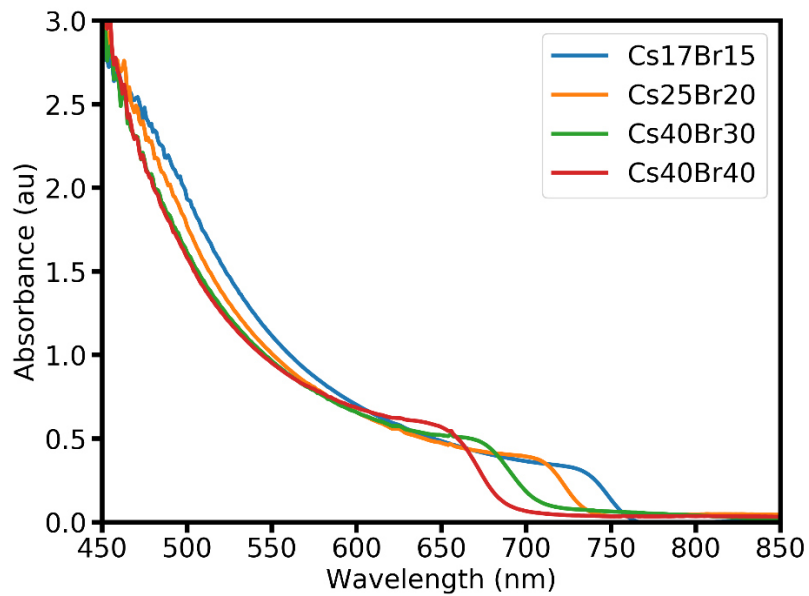

**Figure S4.** UV-Vis absorption spectra of Cs17Br15 ($E_g$ ~1.63 eV), Cs25Br20 ($E_g$ ~1.7 eV), Cs40Br30 ($E_g$ ~1.75 eV) and Cs40Br40 ($E_g$ ~1.8 eV) perovskite thin films showing sharp absorption onsets. Tauc plot from the absorption spectra are used to determine the bandgap.

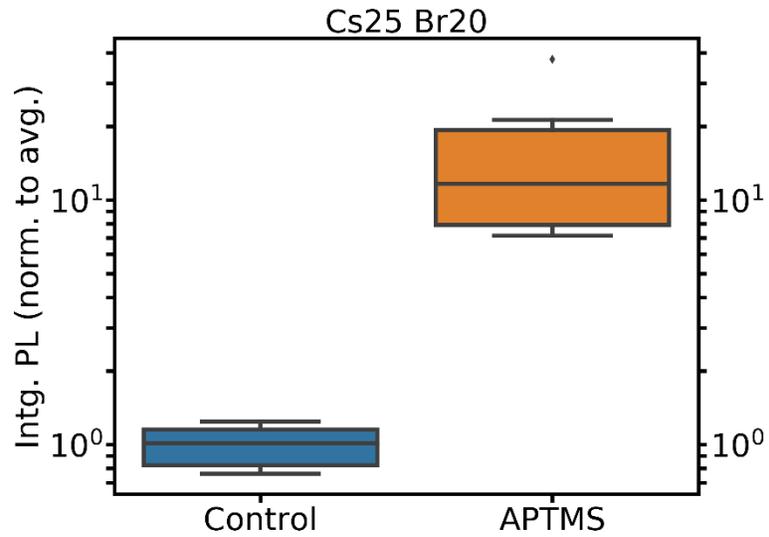

**Figure S5.** Integrated PL intensity normalized to the average for control and APTMS passivated Cs25Br20 films.

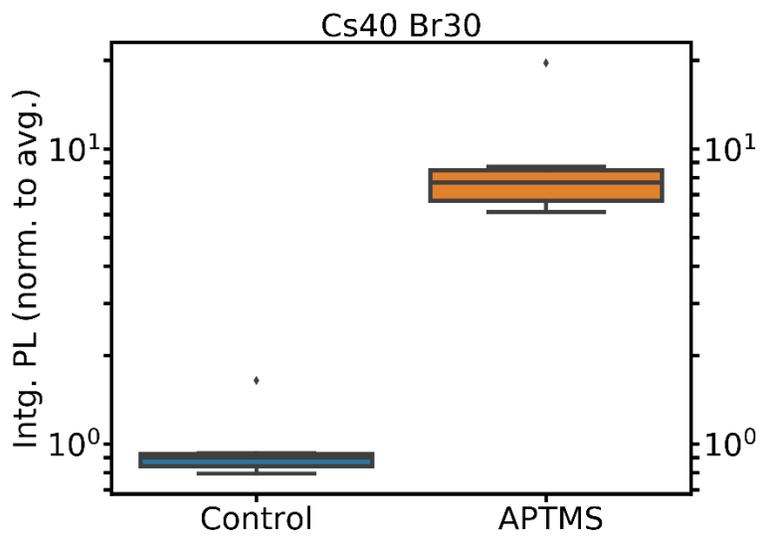

**Figure S6.** Integrated PL intensity normalized to the average for control and APTMS passivated Cs40Br30 films.

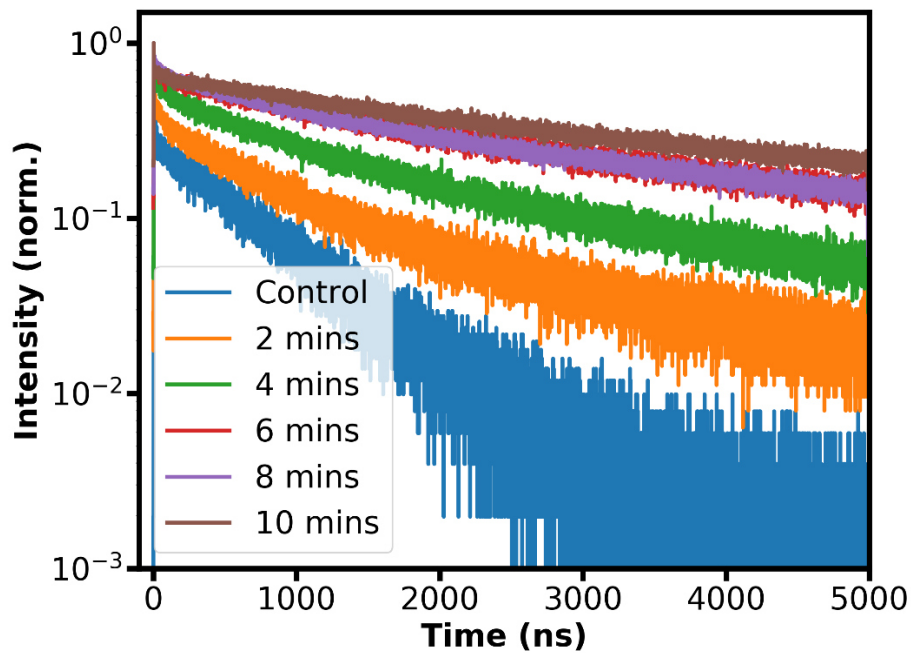

**Figure S7.** Normalized time-resolved PL intensity for control and APTMS passivated Cs17Br15 films as a function of exposure time to APTMS.

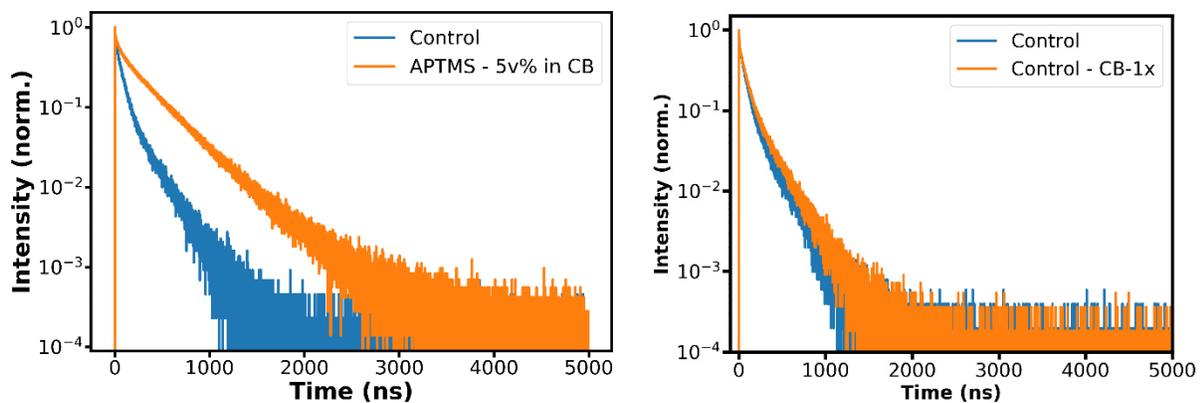

**Figure S8.** Normalized time-resolved PL intensity for (left) control and APTMS passivated Cs17Br15 films by spincoating 5v% APTMS in CB, and for (right) control and only CB spincoated Cs17Br15 films. Together these figures demonstrate that the APTMS is acting as a surface modifier.

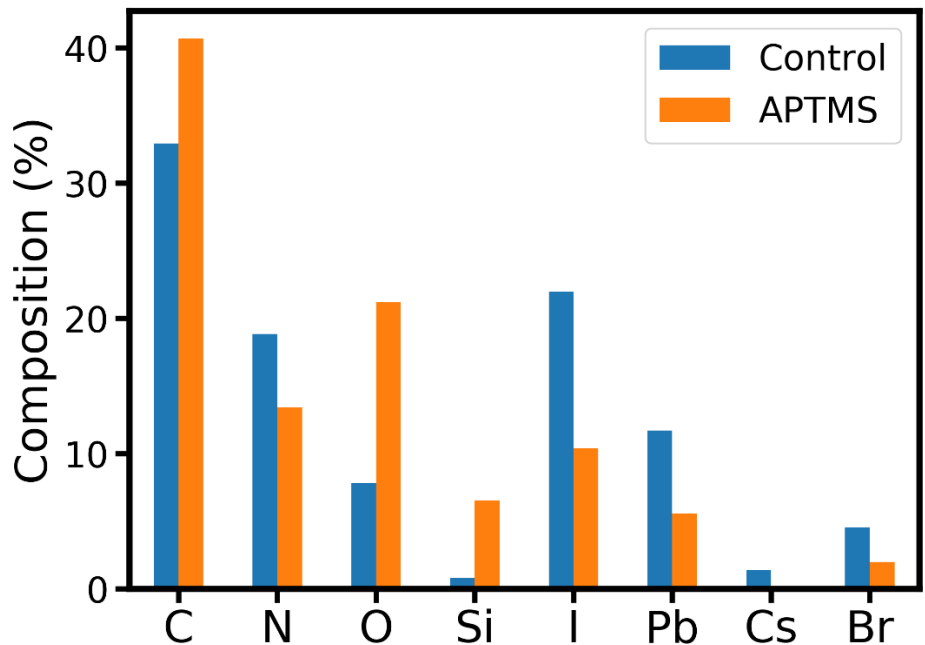

**Figure S9.** Composition determined from XPS survey spectra for control and APTMS passivated Cs17Br15 films. The increase in Si and O concentration demonstrates the presence of APTMS on the perovskite surface. The O 1s signal observed in control samples is adventitious O likely adsorbed on the surface.

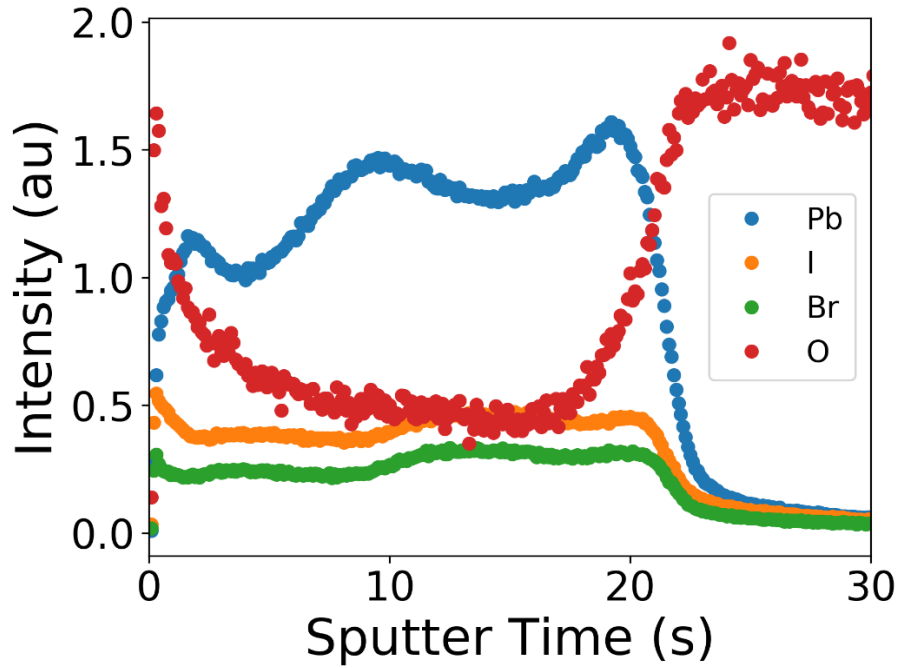

**Figure S10.** Elemental depth profile of silane passivated Cs17Br15 films acquired usign GDOES (glow discharge optical emission spectroscopy) showing decreasing O signal and increasing perovskite signal (Pb, I, Br) as a function of depth (sputtering time). As the perovskite is completely etched away as determined from the decreasing Pb, I, Br signals, we see the O signal rise again, now from the glass substrate underneath the perovskite.

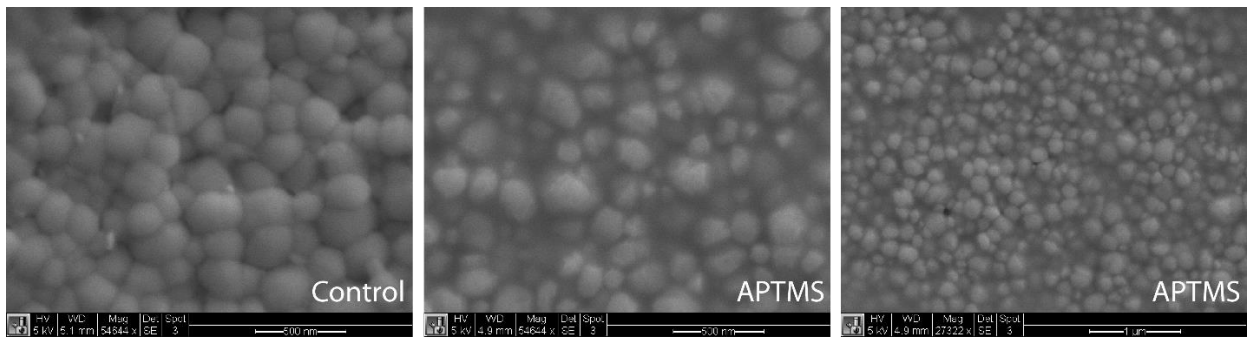

**Figure S11.** SEM morphology of control and APTMS modified Cs17Br15 thin films (on ITO) showing no morphology change of the underlying perovskite after APTMS modification.

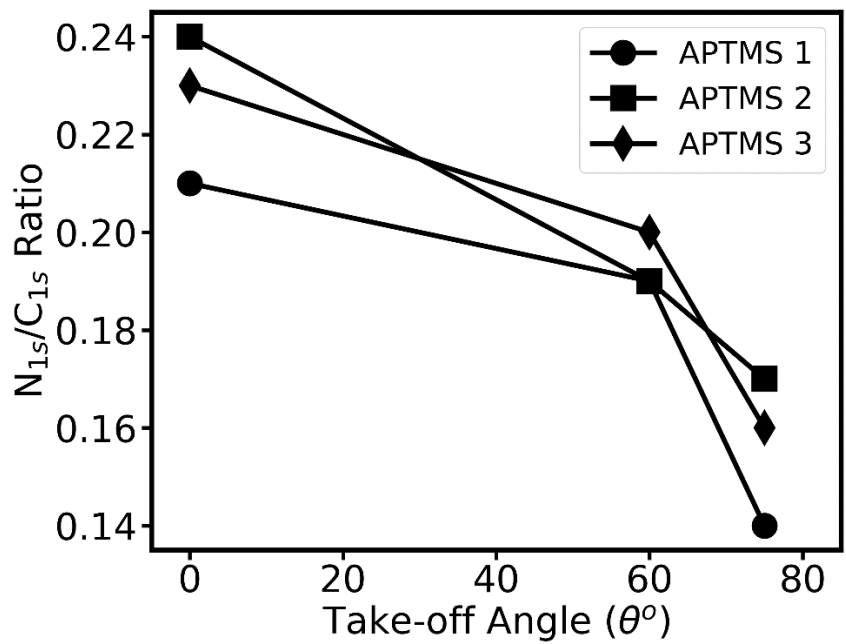

**Figure S12.** Angle-resolved XPS for APTMS passivated film. N 1s/C 1s signal as a function of photo-electron take-off angle. The decreasing N 1s/C 1s signal with increasing take-off angle (more surface sensitivity) indicates that the terminal amine in APTMS is oriented towards the perovskite surface.

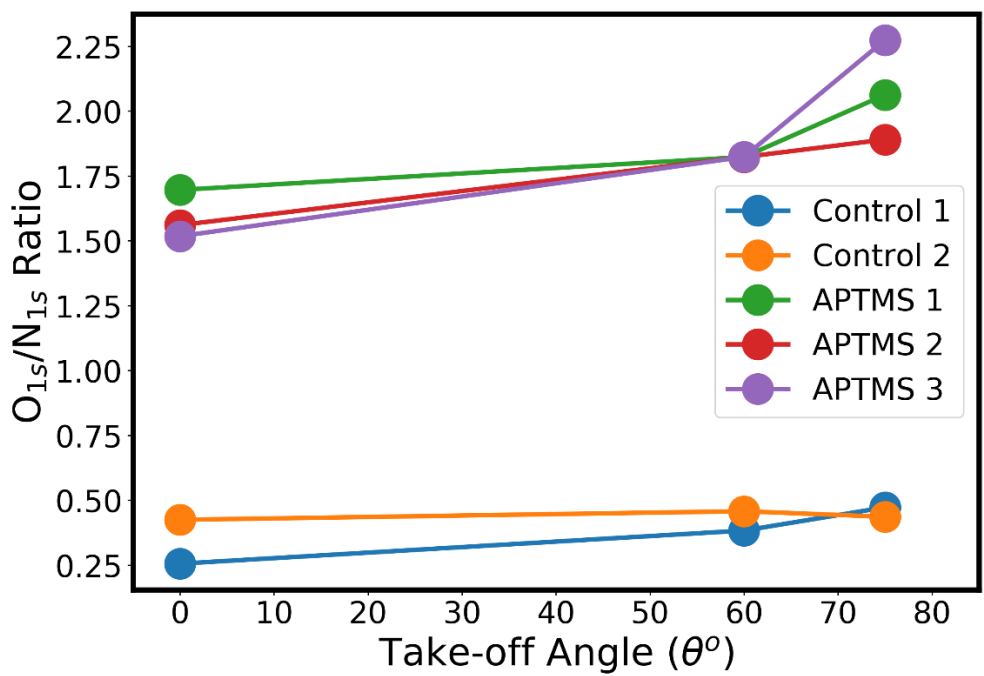

**Figure S13** O 1s/N 1s signal ratio for control and APTMS passivated Cs17Br15 thin films as a function of photo-electron take-off angle. The APTMS samples show an increasing trend of O

1s/N 1s ratio with increasing surface sensitivity, indicating that the methoxy groups on average are oriented away from the perovskite surface. The control samples show no distinct trend. The O 1s signal observed in control samples is adventitious O likely adsorbed on the surface.

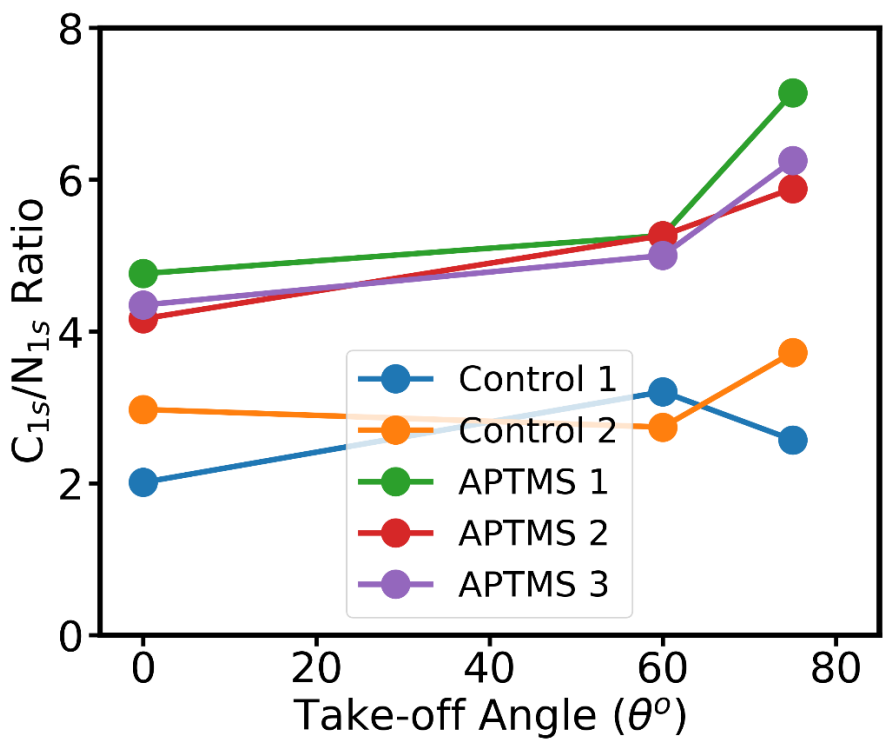

Figure S14. C 1s/N 1s signal ratio for control and APTMS passivated Cs17Br15 thin films as a function of photo-electron take-off angle. The APTMS samples show a distinct trend with increasing C 1s/N 1s ratio with increasing surface sensitivity, indicating that the terminal amine is oriented towards the perovskite surface. The control samples show no distinct trend.

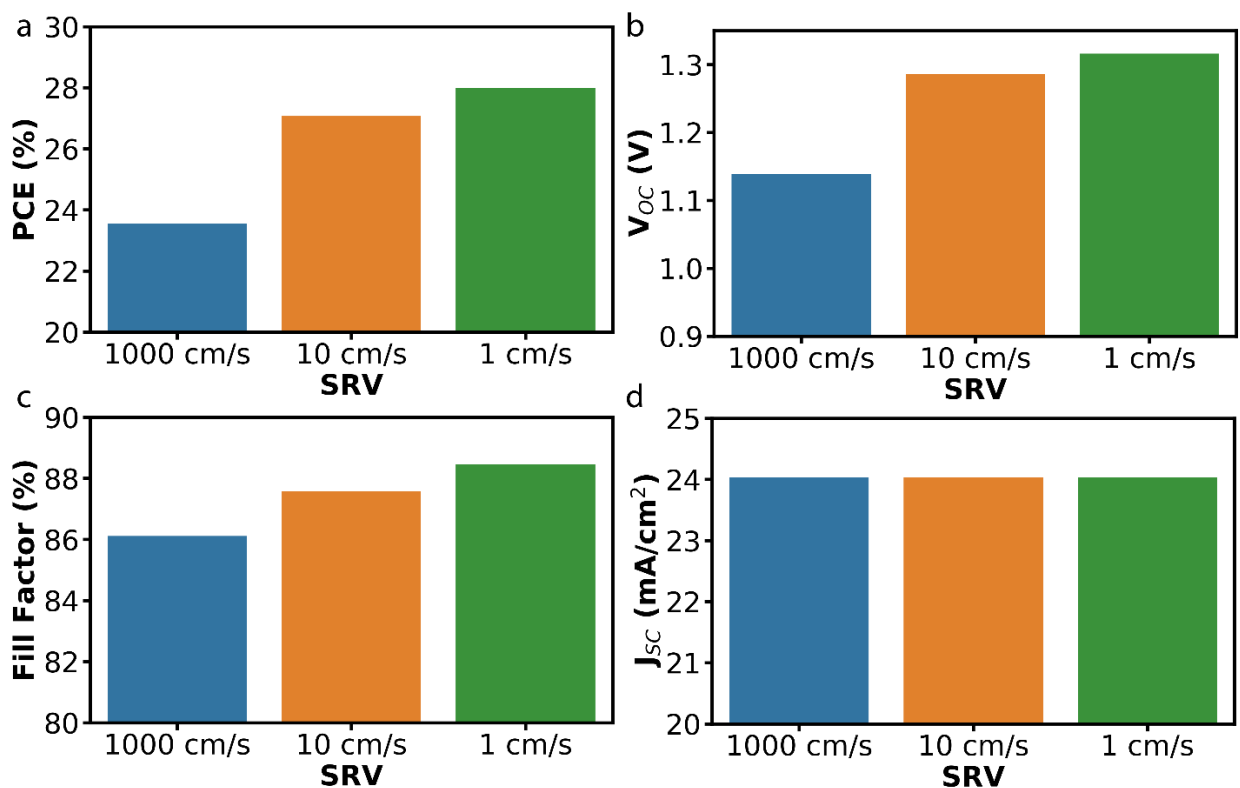

Figure S15. (a) Power conversion efficiency (PCE), (b) open circuit voltage ($V_{OC}$), (c) fill factor and (d) short circuit current density ($J_{SC}$) for J-V SCAPS[6] simulations shown in Figure 4 in the main manuscript.

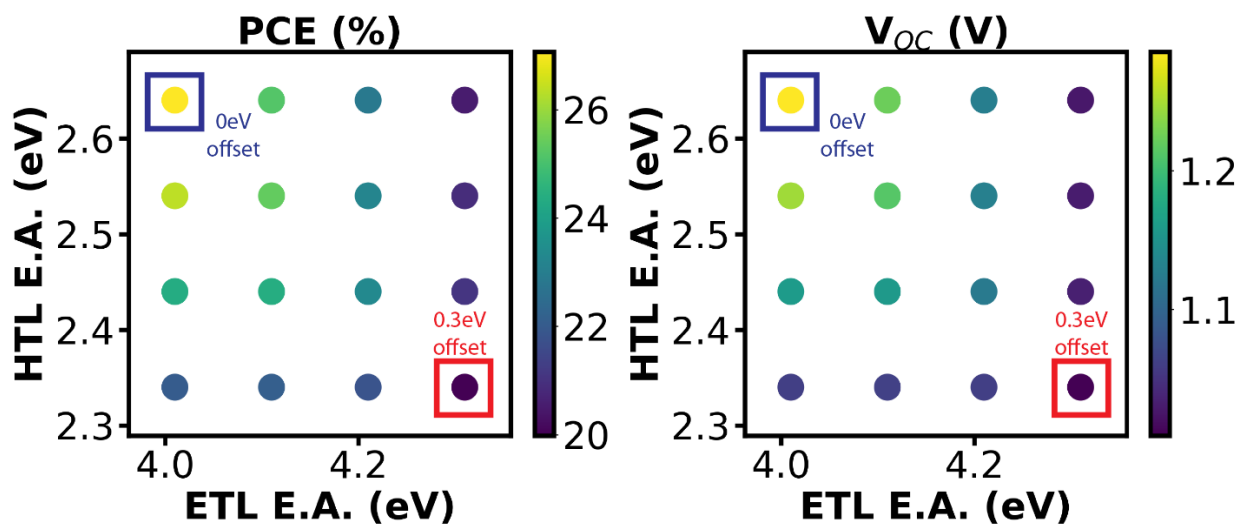

Figure S16. PCE and $V_{OC}$ as a function of different transport layer energy alignments (SRV1=SRV2 = 10 cm/s). Increasing energy offsets decreases the highest achievable PCE and $V_{OC}$. The simulations were performed using SCAPS.[6]